\documentclass[twoside,fleqn]{article}
\usepackage{epsfig,espcrc2}
\newcommand{\nc}{{\cal N}}

\newcommand{\AmS}{{\protect\the\textfont2
  A\kern-.1667em\lower.5ex\hbox{M}\kern-.125emS}}

\title{Measuring an entropy in heavy ion collisions}

\author{A. Bialas, W. Czyz and J. Wosiek\address{M. Smoluchowski Institute of Physics, 
        Jagellonian University, Cracow}%
        \thanks{e-mail: wosiek@thrisc.if.uj.edu.pl} }
        
\begin{document}

\begin{abstract}
We propose to use the coincidence method of Ma to measure an entropy
of the system created in heavy ion collisions. Moreover we estimate,
in a simple model, the 
values of parameters for which the thermodynamical behaviour sets in.
\end{abstract}

\maketitle
%
%
Existence of thermodynamical equilibrium in heavy ion collisions
is an important question. Many phenomenological models critically depend
on this assumption, and consequently there exists an ongoing theoretical
debate of this problem. Alternatively one should develop
 methods which would allow to answer this question directly by 
the exeriments. For example how to check the
saddle point relation
\begin{equation}
\left. \frac{\partial S(E,n)}{\partial E}\right|_{n} = \frac{1}{T},
   \label{1.1}
\end{equation}
which is central in a thermodynamical description. Out of four 
observables entering above equation (energy $E$, multiplicity $n$
 and temperature $T$), the entropy $S$ is most difficult
to directly measure in experiment.
In this talk the recent proposal to determine entropy experimentally
is discussed \cite{WE}. 

The Boltzmann relation $S(E,n)=\log{\Gamma(E,n)}$ reduces the problem to
measuring the density of states $\Gamma(E,n)$ of a system with given
 energy and multiplicity. To this end we have proposed to employ
 the coincidence method advocated by Ma \cite{MA}
\footnote{A simple variant of this technique is used
in common life by fishermen to estimate the nubmer of fish in a lake.
We thank K. Zalewski for this comment.}.  
Suppose that the phase-space is divided into multidimensional cells.
 Suppose furthermore that our
system occupies $\Gamma$ cells (with a uniform probability). Each cell represents a
different state of the system. Our problem is to calculate
$\Gamma$. Let us select randomly
$\nc$ configurations of the system. The main advantage of this approach
is that $\nc\sim\sqrt{\Gamma} \ll \Gamma$ is sufficient.
 These configurations occupy some cells. 
The average occupation number of a cell is $\nc/\Gamma\ll 1$. 
Under this condition, the average number of pairs in 
the same cell 
is
$
(\nc/\Gamma)^2
$
 The total
number of coincidences $N_c$ is the sum over all cells
$
 N_c=\Gamma (\nc/\Gamma)^2,         
$
hence
\begin{equation}
 \Gamma =\frac{N_t}{N_c},            \label{2.1c}
\end{equation}
where $N_t\approx \nc^2$ is the total number of pairs. 
 Generalization for nonequivalent configurations
is also available \cite{WE,bg}.

To define meaningfully a coincidence of two states, we discretize
particle momenta $p_i=a\; n_i$ with some discretization scale $a$.
As usual $a$ should be chosen such that computations are practically 
feasible and at the same time reproduce continuum physics. 

%
We have tested the method in the case of the classical gas of noninteracing,
 nonrelativistic particles with mass $m$ in $d$ dimensions. 
%
 The discretized expression for 
the number of momentum states of $N$ degrees of freedom with the total energy
 $E=M a^2/ 2 m$ reads
\begin{equation}
\Gamma(M,N)=\sum_{n_1,\dots n_N, n_1^2+\dots n_N^2=M} 1,
\end{equation}
where the momenta $p_i=a n_i$ as above.

Numbers $\Gamma(M,N)$ satisfy simple recursion relation which will be used to
calculate them exactly. These will serve to
benchmark the performance of our Monte Carlo.
%

Two questions were addressed: a) how large numbers $\Gamma$ can be reproduced by the
coincidence method with present computers and b) how well is {\em the thermodynamic}
 limit, $M,N \rightarrow \infty$, $M/N=\epsilon$-fixed, 
\begin{equation}
 \frac{1}{N}\log{\Gamma(M,N)}\cong \frac{1}{2}[\log{(\epsilon)}+\log{(2\pi)}+1].
 \label{thermo}
\end{equation}
approximated within available window of $M$ and $N$. 
%

A sample of runs is summarized in Table 1. Instead of generating $\nc$ 
configurations of $N$ integer-valued momenta
$\{n_1,n_2,\dots,n_N\}_k$, $k=1,...,\nc$.
we have uniquelly labelled each, $k$-th say, configuration by an integer
index $I_k$, $k=1,..,\Gamma(M,N)$. Consequently each Monte Carlo run consisted of 
a generation of a sample 
of $\nc$ 
 integer indices, $(I_1,I_2,{\dots},I_{\nc}),
  1\leq I_k \leq \Gamma(M,N), k=1,\dots ,\nc$, uniformly distributed in the whole
space of available states. Then we counted the number of coincidences 
$\hat{N}_c$, i.e., the number of pairs $(I_j,I_k)$ such that $I_j=I_k$. 
The estimate for the number
of all states (column 5 of Table 1) is then
\begin{equation}
\hat{\Gamma}=\nc(\nc -1)/\hat{N}_c.  \label{esti}
\end{equation}
Multinomial nature of the above process allows for simple calculation of the 
distribution of $N_c$ 
In particular, the dispersion of $N_c$ reads ($\nc>>1$) 
\begin{equation}
\sigma^2[N_c]=2<N_c>(1+2\frac{\nc}{\Gamma}+\frac{\nc^2}{2\Gamma^2}),  \label{sigma}
\end{equation}
with $<N_c>=\nc^2/\Gamma$, cf.(\ref{esti}).
This gives for the relative error of the determination of $\Gamma$ after $\nc$ trials
$
\sqrt{\sigma^2[\Gamma]}/\Gamma=\sqrt{2\Gamma}/\nc 
$
and for the estimate of the error 
$
\hat{\sigma}[\Gamma]/\hat{\Gamma}=\sqrt{2/\hat{N}_c}.
$
The last estimate is quoted in column 6-th while the actual relative deviation is in
the last column. 
It is evident from these formulas that the coincidence method works
for much smaller number of trials $(\sim \sqrt{\Gamma})$ than the standard approach
 which measures average occupation of a single state. 
%
%
 The estimated error is steadily
decreasing like $1/\nc$ and actual deviation follows the suit albeit with some fluctuations.
 It is interesting to note that the errors decrease as a 
number
of trials and not as $1/\sqrt{\nc}$, see \cite{WE} for more details.
Substantial improvement in the performance can be achieved if naive
counting of pairs is replaced by "binwise" counting, i.e. a set of
generated indeces $\{I_1,\dots ,I_{\nc}\}$ is first sorted. This trick
reduces the computing effort from $O(\nc^2)$ to $O(\nc\log{\nc})$. 
We find that the Monte Carlo results are well under contol
and show that the method is quite reliable. With the sorting trick it is practical
for $\Gamma$ of the order of $10^{10}$.
 We will discuss now if 
this is sufficient to see the onset of thermodynamic properties.

\begin{table*}[hbt]
\setlength{\tabcolsep}{1.5pc}
\newlength{\digitwidth} \settowidth{\digitwidth}{\rm 0}
\catcode`?=\active \def?{\kern\digitwidth}
\begin{tabular*}{\textwidth}{@{}
l@{\extracolsep{\fill}}
rrrrr}
\hline\hline 
      $N\;\;\;M$    &
      $\nc$         &
      $\hat{N}_c$   &
      $\hat{\Gamma}$&
      $\hat{\sigma}/\hat{\Gamma}$&
      $\delta/\Gamma$             \\  
\hline
                     &
      $ 800\; 000$     & 
      $ 10     $     &
      $ 6.39999_{10}+10$  & 
      $  0.44$       & 
      $  0.24  $          \\
      $31\;\;\;16$   & 
      $ 1\; 600\; 000$   & 
      $ 42 $         &
      $ 6.09523_{10}+10$   &
      $ 0.22 $       &
      $ 0.18 $            \\
                     & 
      $3\; 000\; 000$    & 
      $188$          &
      $ 4.78723_{10}+10$  & 
      $ 0.10$        &
      $ 0.08   $           \\
\hline
$\Gamma$
                      & 
                      &
                      &
      $51\; 795\; 303\; 424$&                  
                      &
                            \\
\hline
                     &
      $ 750\; 000$     & 
      $ 12     $     &
      $ 4.68749_{10}+10$  & 
      $  0.40$       & 
      $  0.13  $          \\
      $13\;\;\;24$   & 
      $ 1\; 500\; 000$   & 
      $ 74$         &
      $ 3.04054_{10}+10$   &
      $ 0.16 $       &
      $ 0.27 $            \\
                     & 
      $3\; 000\; 000$    & 
      $256$          &
      $ 3.51562_{10}+10$  & 
      $ 0.09$        &
      $ 0.15   $           \\
\hline
$\Gamma$
                      & 
                      &
                      &
      $41\; 469\; 483\; 552$&                  
                      &     \\
\hline\hline
\end{tabular*}
\caption{Monte Carlo results for $\hat{\Gamma}(M,N)$ (col.5)
.
 The third and fourth column
 give the number of generated configurations $\nc$, and the number of observed
coincidences $\hat{N}_c$. In the last two columns we quote
the Monte Carlo estimate of the relative error, see the text, and the 
actual relative deviation $\delta/\Gamma=|\hat{\Gamma}-\Gamma|/\Gamma$
 from the exact value $\Gamma$ also quoted in the Table.}

\end{table*}
%

     Figure 1 shows the entropy density as a function of the 
scaling variable $\epsilon=M/N$.
Statistical errors of MC results (and the deviation from the exact discrete values given by 
$\Gamma(M,N)$) are much smaller that the size of symbols. Exact values are very close 
to the continuum expressions for the volume of the nonrelativistic phase space (solid
lines).
 Considered as a 
function of $\epsilon$ and $N$ they obviously show a substantial $N$-dependence. 
The $N$ varies from 8 (lowest curve) to 24 in
 this plot. On the other hand, the deviation from the ultimate scaling limit, 
(Eq.(\ref{thermo}), the uppermost curve), is around 30\% in the worst case (N=8,M=30).
With $N$ starting from 12, deviations from the infinite system are smaller than 20\%.
Note that $N$ denotes the number of degrees of freedom, which in $d$ space dimensions 
 corresponds to $N/d$ particles.

\begin{figure}[htb]
\vspace{9pt}
\epsfig{width=7.2cm,file=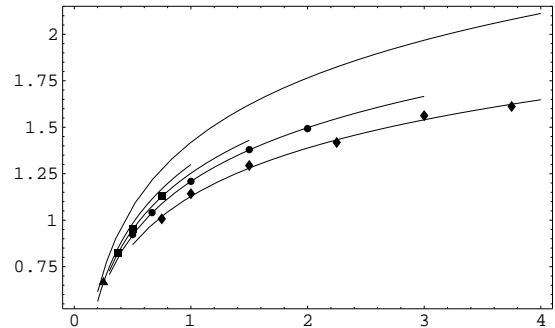}
\caption{Entropy density $s=\frac{1}{N}\log{\Gamma(M,N)}$ vs.  
 $\epsilon$. Black symbols represent our Monte Carlo results for N=8 (diamonds),
12 (circles), 16 (boxes) and 24 (a triangle). 
}
\label{fig:largenenough}
\end{figure}
 Second, we test the saddle point relation (\ref{1.1})
with the temperature eliminated with the aid of the equipartition
relation $E/N=T/2$. Upon discretization it reads
\begin{equation}
\log{\left(\frac{\Gamma(M+1,N)}{\Gamma(M,N)}\right)}=\frac{N}{2M+1}. \label{dsde}
\end{equation}

This equation is tested in Fig.2, where a half of the inverse of the left hand side,
as obtained from simulations,
is plotted as a function of $\epsilon$. Solid line represents the right hand side 
\footnote{Of course $\epsilon=(M+1/2)/N$ in this case.}.
Similarly to the previous case agreement is very good for $N\geq 12$. It was necessary
to reduce MC errors to the level of 1\%-3\% in order to achieve this agreement. Of couse
this test is much more sensitive than the previous one since it requires precise
 measurement of the derivatives.

\begin{figure}[htb]
\epsfig{width=7.2cm,file=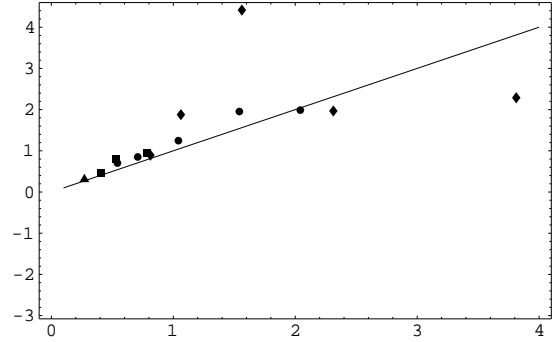}
\caption{Testing the relation \protect{(\ref{dsde})}.
Symbols as in Fig.1, solid line corresponds to the
thermodynamical lmit.}
\label{fig:toosmall}
\end{figure}

To conclude, the coincidence method is satisfactory in practice for the number of degrees
of freedom (or number of cells) up to $\sim 30$. This is sufficient to see 
the signatures
of the thermal equilibrium. For more than 12 degrees of fredom the scaling of the
 entropy density is confirmed with the
 accuracy better than 20\% . The saddle point relation, coupled
with the equipartition principle, $\partial S/\partial E=1/2\epsilon$
 is also very well reproduced. 

\vspace*{.5cm}
This work is supported in part by the Polish Committee for Scientific
Research under the grants no. 2P03B 08614 and 2P03B 04412.

\end{document}